\documentclass[10pt,conference]{IEEEtran}

\IEEEoverridecommandlockouts
\usepackage{cite}
\usepackage{amsmath,amssymb,amsfonts}
\usepackage[ruled,linesnumbered]{algorithm2e}
\usepackage{algorithmic}
\algsetup{linenosize=\small}
\SetKwInput{KwInput}{Input}                
\SetKwInput{KwOutput}{Output}              
\usepackage{graphicx}
\usepackage{textcomp}
\usepackage{xcolor}
\usepackage{enumitem}
\usepackage{url}
\usepackage{tcolorbox}
\usepackage{balance}
\usepackage{tcolorbox}
\usepackage[multiple]{footmisc}
\usepackage[english]{babel}
\usepackage{csquotes}
\usepackage{multirow}
\usepackage{subfigure}
\usepackage{lipsum}
\usepackage{tikz}

\usepackage{xcolor}
\usepackage{enumitem}
\usepackage{amsmath,amssymb}

\usepackage{comment}
\newboolean{showcomments}
\setboolean{showcomments}{true}
\ifthenelse{\boolean{showcomments}}
{ }

\ifthenelse{\boolean{showcomments}}
{ }

\ifthenelse{\boolean{showcomments}}
{ }

\def\BibTeX{{\rm B\kern-.05em{\sc i\kern-.025em b}\kern-.08em
    T\kern-.1667em\lower.7ex\hbox{E}\kern-.125emX}}
\begin{document}

\newcommand{\yz}[1]{\textcolor{cyan}{{#1}}}
\newcommand{\jh}[1]{\textcolor{red}{\textbf{JH:} #1}}

\title{What Do Users Ask in Open-Source AI Repositories? An Empirical Study of GitHub Issues}

\author{
	\IEEEauthorblockN{Zhou Yang\IEEEauthorrefmark{1},
	Chenyu Wang\IEEEauthorrefmark{1},
	Jieke Shi\IEEEauthorrefmark{1},
	Thong Hoang\IEEEauthorrefmark{2},
	Pavneet Kochhar\IEEEauthorrefmark{3},
	Qinghua Lu\IEEEauthorrefmark{2},
	Zhenchang Xing\IEEEauthorrefmark{2}\IEEEauthorrefmark{4},
	David Lo\IEEEauthorrefmark{1}}
	\IEEEauthorblockA{\IEEEauthorrefmark{1}Singapore Management University, Singapore}
	\IEEEauthorblockA{\IEEEauthorrefmark{2}Data61, Eveleigh, CSIRO, Sydney, Australia}
	\IEEEauthorblockA{\IEEEauthorrefmark{3}Microsoft Research, Vancouver, Canada}
	\IEEEauthorblockA{\IEEEauthorrefmark{4}Australian National University, Canberra, Australia}
	\IEEEauthorblockA{\{zyang, chenyuwang, jkshi, davidlo\}@smu.edu.sg}
	\IEEEauthorblockA{\{James.Hoang, Qinghua.Lu, Zhenchang.Xing\}@data61.csiro.au}
	\IEEEauthorblockA{Pavneet.Kochhar@microsoft.com}
}

\maketitle

\begin{abstract}

Artificial Intelligence (AI) systems, which benefit from the availability of large-scale datasets and increasing computational power, have become effective solutions to various critical tasks, such as natural language understanding, speech recognition, and image processing. The advancement of these AI systems is inseparable from open-source software (OSS). Specifically, many benchmarks, implementations, and frameworks for constructing AI systems are made open source and accessible to the public, allowing researchers and practitioners to reproduce the reported results and broaden the application of AI systems. The development of AI systems follows a data-driven paradigm and is sensitive to hyperparameter settings and data separation. Developers may encounter unique problems when employing open-source AI repositories.

This paper presents an empirical study that investigates the issues in the repositories of open-source AI repositories to assist developers in understanding problems during the process of employing AI systems. We collect 576 repositories from the \texttt{PapersWithCode} platform. Among these repositories, we find 24,953 issues by utilizing GitHub REST APIs. Our empirical study includes three phases. First, we manually analyze these issues to categorize the problems that developers are likely to encounter in open-source AI repositories. Specifically, we provide a taxonomy of 13 categories related to AI systems. The two most common issues are runtime errors (23.18\%) and unclear instructions (19.53\%). Second, we see that 67.5\% of issues are closed. We also find that half of these issues resolve within four days. Moreover, issue management features, e.g., label and assign, are not widely adopted in open-source AI repositories. In particular, only 7.81\% and 5.9\% of repositories label issues and assign these issues to assignees, respectively. Finally, we empirically show that employing GitHub issue management features and writing issues with detailed descriptions facilitate the resolution of issues. Based on our findings, we make recommendations for developers to help better manage the issues of open-source AI repositories and improve their quality.

\end{abstract}


\section{Introduction}
\label{sec:intro}

Artificial Intelligence (AI) has revolutionized many critical tasks in our modern lives, such as natural language understanding~\cite{danilevsky2020survey, chowdhary2020natural}, image processing~\cite{tjoa2020survey}, speech recognition~\cite{HuBERT}, and code understanding~\cite{9825884,alsulami_source_2017}.
The success of these AI techniques is heavily dependent on the open-source culture. 
Popular open-source repositories like \texttt{TensorFlow}\footnote{\url{https://www.tensorflow.org/}} and \texttt{PyTorch}\footnote{\url{https://pytorch.org/}} have greatly facilitated the development of AI applications. 
Additionally, research papers published at flagship AI conferences often release their replication packages and models, making it easier for researchers and practitioners to replicate results, run models on other tasks, or conduct further research. Furthermore, open-source benchmarks such as CodeXGLUE\footnote{\url{https://github.com/microsoft/CodeXGLUE}} and ImageNet\footnote{\url{https://www.image-net.org/}} have played a crucial role in the advancement of AI by providing a common platform for researchers to evaluate their models and compare their results.

Despite the benefits of open-source AI repositories, developers often encounter challenges when utilizing them to build AI applications. A recent article~\cite{heaven_2022} points out that many researchers have complained about a `replication crisis' in AI repositories. This issue can be attributed to various factors, such as incomplete instructions for configuring development environments~\cite{gundersen2018reproducible} and a lack of information about the datasets used to train AI models. Additionally, buggy code can also be a significant obstacle for developers. These problems can lead to frustration and hinder the use of open-source AI repositories.

To address the problems above, developers should communicate with repository owners and other developers to discuss and address these problems together. There are several channels for such communication. 
For example, developers can post questions in StackOverflow\footnote{\url{https://stackoverflow.com/}} (SO). However, SO is unsuitable for finding answers to questions specific to a repository as there may be fewer experts on this topic. Another channel is email. Researchers or the repositories' owners usually publish their contact information, e.g., email addresses and affiliations, in their papers or the descriptions of their repositories, respectively. The developers can also inquire about questions via email; however, this communication is often private and inaccessible for analysis. The issue trackers, such as GitHub Issues\footnote{\url{https://github.com/features/issues}} or JIRA,\footnote{\url{https://www.atlassian.com/jira}} can be appropriate resources to analyze the problems faced by the AI repository users.

We obtain a list of papers published between 2013 and 2022 from ten top-tier AI conferences and retrieve their replication packages from the \texttt{PapersWithCode}\footnote{\url{https://paperswithcode.com/}} platform. After cleaning the data (e.g., removing the repositories that are not provided by the authors of papers), we identify 576 open-source software (OSS) AI repositories hosted on GitHub and 24,953 GitHub issues using the GitHub REST APIs. We then conduct open card sorting to develop a taxonomy of the issues in open-source AI repositories, categorizing them into 13 categories. Our findings reveal that the most commonly reported issue is \texttt{Runtime Error}.
The second-largest group of issues falls under \texttt{Unclear Instructions}, primarily caused by inadequate documentation.

Our study examines how developers manage and address issues in their open-source AI repositories. The results indicate that 67.53\% of issues are resolved, with 50\% of issues closed within four days. Additionally, we found a positive correlation between the number of issues in a repository and the average time to address the issues (i.e., the time between the open and close date of issues), as well as a negative correlation between the number of issues and the closed issue rate (i.e., the ratio of closed issues to all the issues in a repository). We also discovered that developers often neglect the features provided by GitHub to manage issues, with only 7.81\% of repositories using labels to categorize issues and only 5.90\% of repositories assigning issues to specific individuals. Further analysis of the relationship between various features and the closure of issues in open-source AI repositories reveals that an issue's label(s) and assignee(s) have a statistically significant impact on the closure of issues. These findings suggest that repository maintainers should actively utilize GitHub issue management tools to effectively manage issues in their repositories. Additionally, issues with longer descriptions and code blocks are more likely to be closed, indicating that issue raisers should provide more details, especially code blocks, to convey information clearly.

We make the following contributions in this paper:
\begin{itemize}
	\item We conduct an empirical study to systematically investigate the issues in open-source AI repositories hosted on GitHub. The study contributes to the community with a dataset of 576 AI repositories that are manually confirmed to be the official implementations of papers from top-tier AI conferences and 24,953 issues in these repositories.
	\item We categorize the issues in open-source AI repositories to help developers better understand the problems users encounter when using these systems.
	\item Our empirical study shows that repository maintainers and developers should follow some good practices to help better address the issues in open-source AI repositories. We encourage the maintainers to actively adopt issue management features, e.g., the labeling and assigning features provided by GitHub. We also suggest that the developers provide detailed information, especially code blocks, and format the issues properly.
\end{itemize}

The rest of this paper is organized as follows. Section~\ref{sec:background} explains the background of open-source AI repositories and their issues.
In Section~\ref{sec:methodology}, we describe the process of collecting and cleaning datasets.
Section~\ref{sec:results} reports research questions and experiment results. 
We discuss the difference between academic and industry repositories, as well as the threats to validity in Section~\ref{sec:discussion}. 
Section~\ref{sec:related_work} reports the works relevant to our work.
We conclude the paper and present future work in Section~\ref{sec:conclusion}.
\section{Background}
\label{sec:background}


\begin{figure*}[!t]
	\centering
	\includegraphics[width=0.9\linewidth]{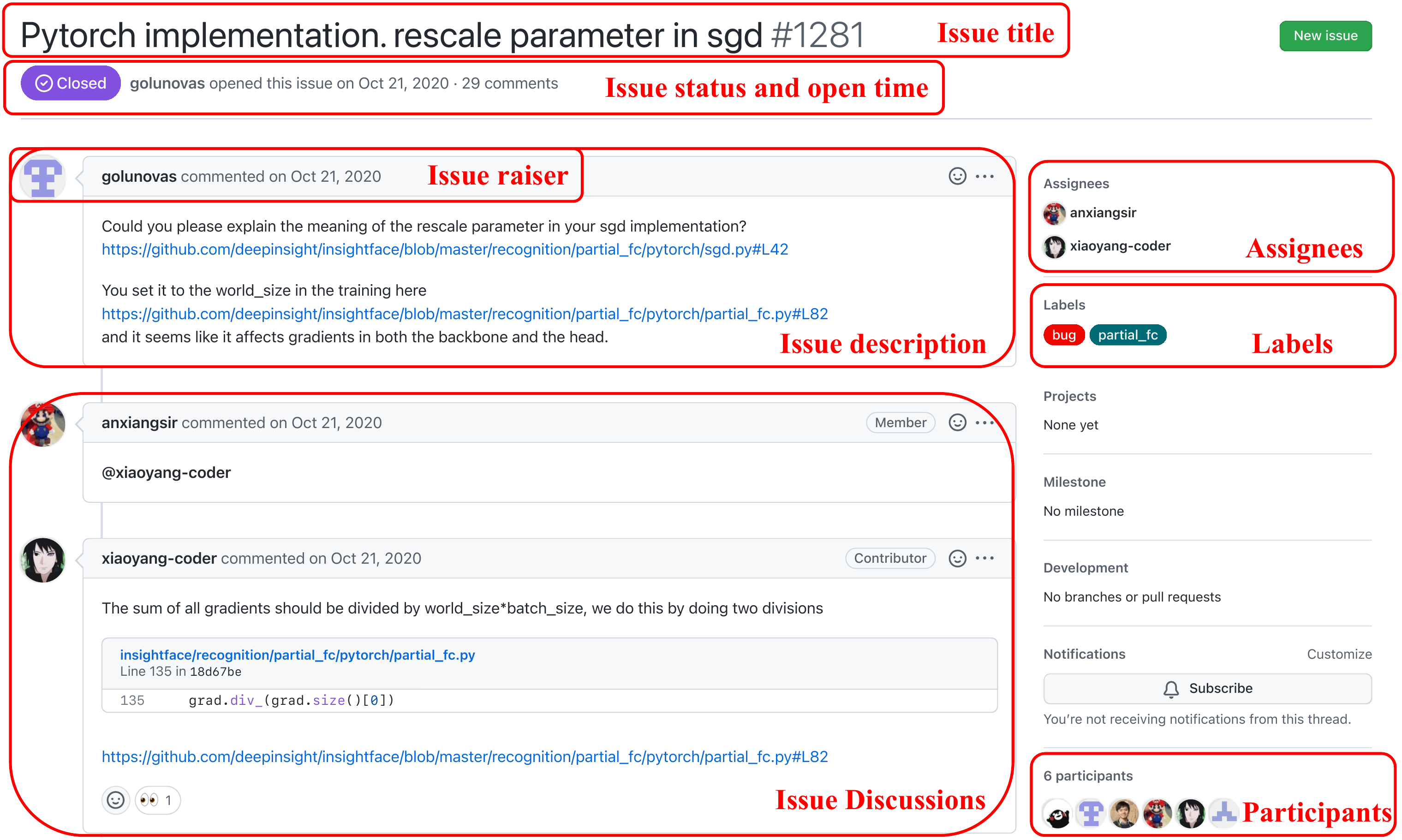}
	\caption{An example of an issue in an AI repository hosted on GitHub.
    }
    \label{fig:issue-example}
\end{figure*}

\subsection{Open-Source AI Repositories}

The continued success of open-source software (OSS) in the field of AI, as exemplified by projects such as Tensorflow, has been widely acknowledged by both practitioners and researchers. The critical role that OSS plays in the advancement and development of AI techniques has been emphasized in various studies~\cite{sonnenburg2007need,Spohrer_2021}. In particular, the following five AI-related components should be made open source: design, implementation, frameworks, benchmarks, and models~\cite{Spohrer_2021}.

Artificial intelligence (AI) papers and technical reports that describe the design of algorithms and models are often made available to the public in the form of open-source materials. For example, authors are encouraged to publish their papers as open-access or release preprint versions on platforms such as \texttt{arXiv}. Additionally, the review process for many AI conferences is conducted publicly on platforms like \texttt{OpenReview}. Furthermore, the implementation of AI systems is commonly made open source as well. Developers release their code on platforms such as GitHub and BitBucket, enabling researchers and practitioners to replicate results and conduct further extensions. The key frameworks that support the implementation of AI software, such as \texttt{Tensorflow} and \texttt{PyTorch}, are also made open source to facilitate the development of various AI models. There are emerging open-source frameworks for specific tasks, such as \texttt{Carla} for autonomous driving~\cite{dosovitskiy2017carla} and \texttt{OpenAI Gym} for reinforcement learning~\cite{brockman2016openai}.


Open-source public benchmark platforms, e.g., \texttt{CodaLab}\footnote{\url{https://codalab.org/}} and \texttt{Kaggle},\footnote{\url{https://www.kaggle.com/}} provide developers with opportunities to participate in competitions and share their AI solutions. Platforms also offer large pre-trained AI models that can be fine-tuned for downstream tasks. For instance, HugginingFace\footnote{\url{https://huggingface.co/}} is a platform that shares a variety of Transformer-based models and collections of datasets for various tasks, including natural language processing, audio processing, and speech recognition. Similarly, ModelZoo\footnote{\url{https://modelzoo.co/}} is another platform that shares pre-trained AI models. Giant tech companies like IBM and Google also actively participate in the open-source community by releasing their AI-related projects~\cite{strickland2019ibm, carvalho2019off, goralski2020artificial, tomavsev2020ai}.
Among these open-source platforms, GitHub is one of the most popular and hosts numerous AI repositories~\cite{zhang2019explorative, gonzalez2020state}. Therefore, in this study, we conduct an analysis of open-source AI repositories on GitHub to gain insights into their usage and popularity.

\subsection{Tracking Issues in GitHub}

The software system is not always reliable~\cite{heaven_2022, gundersen2018reproducible}. Developers of open-source software (OSS) repositories may encounter difficulties when employing these repositories. Introduced in 2009 for the first time, GitHub Issues~\cite{preston-werner_2009} is a built-in issue-tracking system integrated into GitHub that allows developers to raise questions, request features, etc.

Figure~\ref{fig:issue-example} illustrates an example of an issue in an open-source AI repository hosted on GitHub.\footnote{\url{https://github.com/deepinsight/insightface/issues/1281}} The person raising the issue, referred to as the \textit{issue raiser}, provides a title and a description of the issue. The issue can be in either an \textit{open} or \textit{closed} status, indicating whether it has been resolved or not. Any GitHub users who can access the repository can provide \textit{comments} to discuss the issue, such as adding extra details or suggesting solutions. The issue tracker has two basic functions for issue management: a \textit{label(s)} and assign issues to an \textit{assignee(s)}. Anyone with write access to the repository can create a label, and anyone with triage access to the repository can apply and dismiss labels. Only those with write access to the repository can specify the assignee(s). All the users involved in an issue, including the issue raiser, the commenter(s), and the assignee(s) are referred to as \textit{participants}. 


There are various challenges when employing open-source AI repositories. For example, missing datasets and descriptions can make it difficult for users to evaluate AI models and replicate their results. Additionally, pre-trained AI models should be provided so that users can fine-tune them for other downstream tasks. Furthermore, frameworks used to develop AI software applications are constantly evolving; therefore, installing compatible development environments can be challenging for new users. Moreover, the documentation of open-source AI repositories may only mention the names of installed packages without specifying their versions, which can lead to runtime errors. Lastly, AI models are sensitive to hyperparameters~\cite{yang2020estimating, zhuang2022randomness}, and attempts to replicate the results may fail when using random hyperparameters. These issues can impede the usability of AI repositories. In this paper, we aim to understand these challenges to enable users to make better use of open-source AI repositories when developing their software applications.
\section{Data Collection}
\label{sec:methodology}

This section presents the process of collecting and cleaning the dataset, i.e., GitHub issues of open-source software AI repositories, for our empirical study. 

\subsection{Scope}

Following a previous study on analyzing open-source software (OSS) AI repositories~\cite{fan2021makes}, we include five prestigious AI conferences in our experiments: NeurIPS, ICML, CVPR, ICCV, and ECCV. Out of these five conferences, CVPR, ICCV, and ECCV are computer vision-oriented. We also collect papers and repositories from other highly regarded AI conferences, such as ICLR, ACL, SIGKDD, AAAI, and AAMAS. 
All the selected conferences are categorized as A* in the CORE Ranking.\footnote{https://www.core.edu.au/conference-portal}
In total, our dataset is constructed from ten AI conferences that cover a wide range of AI topics, including natural language processing, reinforcement learning, etc. It is worth noting that frameworks and applications of AI techniques are constantly evolving. Our dataset includes repositories related to AI conferences from 2013 to 2022.

\subsection{Data Curation}

\begin{figure}[!t]
	\centering
	\includegraphics[width=1\linewidth]{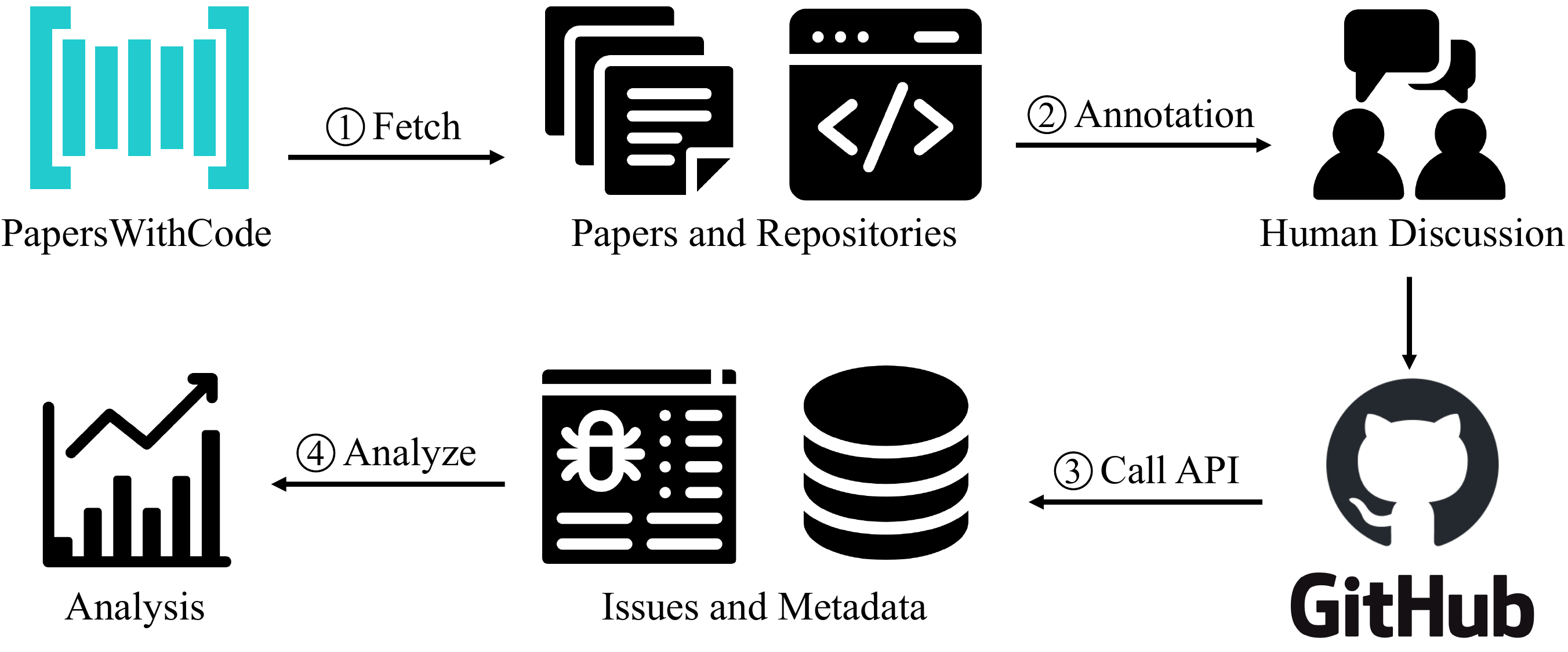}
	\caption{The workflow of collecting and cleaning the data. We obtain the paper list and the corresponding repository list from \texttt{PapersWithCode}. After manually removing the unofficial repositories, we query the GitHub API to obtain the metadata and issue information.}
	\label{fig:workflow}
\end{figure}

Figure~\ref{fig:workflow} illustrates an overview of the workflow for collecting and cleaning our dataset. We use \texttt{PapersWithCode}\footnote{\url{https://paperswithcode.com}} (\texttt{PwC}), a platform for storing AI-related papers, implementations, and datasets, to obtain AI papers published at conferences and their corresponding repositories. According to the contributing guidelines, \texttt{PwC} users can connect a repository to a paper by adding GitHub, GitLab, or BitBucket URLs. Furthermore, users can monitor all edits on a Slack\footnote{\url{https://slack.com/intl/en-au/}} channel, where everyone is  encouraged to review their contributions to ensure data quality. If contributors acknowledge that a repository is the implementation provided by the authors of the corresponding paper, they mark this repository as \texttt{official}.

In this paper, we adopt a conservative strategy for selecting repositories. Specifically, we only consider papers whose linked repositories are labeled as \texttt{official}. Intuitively, users tend to ask questions in the official repository maintained by the authors, and such official repositories are more likely to contain more issues from which more actionable suggestions can be derived.
We use \texttt{PwC} APIs\footnote{\url{https://paperswithcode.com/api/v1/docs/}} to obtain papers from different venues and their metadata, including a list of authors and their repository links. We then conduct a series of queries from the \texttt{PwC} APIs with the following constraints:
\begin{itemize}
	\item \texttt{Venue}: Papers from the ten selected conferences. 
	\item \texttt{Year}: The dates of these conferences are between 1/1/2013 and 31/12/2022.
	\item \texttt{Authority}: All the repositories that are labeled as \texttt{official} on \texttt{PapersWithCode}.
\end{itemize}

In our initial collection, we obtain 652 open-source AI repositories and their corresponding papers from \texttt{PapersWithCode}. Among these repositories, 639 (98.01\%) are published on GitHub, nine (1.38\%) are hosted on CodaLab, and only four (0.61\%) are hosted on Bitbucket. Given the prevalence of GitHub as a platform for hosting open-source AI repositories and the convenience of utilizing GitHub APIs, we decided to focus on the repositories hosted on GitHub for our analysis, resulting in a set of 639 open-source AI repositories.

Subsequently, we use the \texttt{git clone} command to download these 639 open-source AI repositories and manually check whether the downloaded repositories are indeed the \texttt{official} implementation of the corresponding papers. To verify the relationship between repositories and their corresponding papers, we utilize three strategies. Firstly, the README.md file of an AI repository should contain the title of the corresponding paper. Secondly, the paper must mention the repository URL. Lastly, the repository owners should be listed as the authors of the corresponding paper. To evaluate the accuracy of these strategies, we asked two annotators to manually examine each pair of a repository and its corresponding paper to evaluate the accuracy of our strategies.
Following prior studies~\cite{biasfinder}, we employ Cohen's Kappa~\cite{mchugh2012interrater} to measure the inter-annotator agreement, which is 0.72. 
According to~\cite{landis1977measurement}, this inter-annotator agreement value indicates substantial agreement among our annotators.
In cases of disagreement, a third annotator was consulted to finalize the decision. The three annotators, who are Ph.D. students with over five years of experience in using GitHub, were able to confirm 576 pairs of repositories and their corresponding papers as \texttt{official} implementations.

We also employ the GitHub REST API\footnote{\url{https://docs.github.com/en/rest?apiVersion=2022-11-28}} to fetch the metadata and issue information of the collected repositories. The metadata of these repositories includes information such as contributors, the number of stars, the number of forks, the number of watchers, the creation date, and the last update date. We also obtain the issue list of each repository and compute the number of open and closed issues. For each issue, we extract its title, discussion history, status, creation date, close date (if the issue has been closed), assignees, and labels.
Our final dataset consists of 576 pairs of repositories and their corresponding papers. Additionally, we find that 24,953 issues belong to these repositories.

\section{Results}
\label{sec:results}

In this section, we present our analysis of the issues in open-source software (OSS) AI repositories. 
We conduct experiments to answer the following three research questions:

\begin{itemize}
	\item \textit{RQ1. What do users discuss in the issues of open-source AI repositories?}
	\item \textit{RQ2. How are issues in open-source AI repositories managed and addressed?}
	\item \textit{RQ3. What are the relationships between different features and the closure of issues in open-source AI repositories?}
\end{itemize}

\subsection{RQ1. What do users discuss in the issues of open-source AI repositories?}

\noindent \textbf{Motivation.}
The development of AI systems follows a data-driven paradigm~\cite{ntoutsi2020bias, wolf2020ai}, in which AI models learn knowledge from a large amount of data automatically, as opposed to traditional software systems where developers encode their programming language through control flow and data flow graphs~\cite{allen1970control, kavi1986formal}. Additionally, AI systems often require specialized hardware, such as a GPU or Google TPU, to accelerate the training and testing processes. Therefore, the issues that developers face in open-source AI repositories may differ from those in traditional OSS repositories. In order to better understand the primary issues encountered in the development and usage of AI applications, we aim to construct a taxonomy of the issues in open-source AI repositories.

\noindent \textbf{Categorization Process.}
We obtain 24,953 issues from the 576 repositories using the GitHub REST API. Among these repositories, 122 (21.18\%) have no GitHub issues. Previous studies~\cite{10.1145/2568225.2568233,10.1145/1718918.1718973} have used the card sorting method, a technique for creating mental models and deriving taxonomies from data, to categorize software engineering-related discussions. As the issues in open-source AI repositories can be unique, we adopt open card sorting, which uses no predefined groups to create the taxonomy for issues in AI repositories. The taxonomy is intended to categorize \textit{the primary reason why developers raise an issue} rather than the \textit{root cause of this issue}, the latter of which is usually difficult to determine without sufficient information. For example, a developer may raise an issue asking why the results are impossible to reproduce, but the root cause can be due to many reasons, such as differences in library versions, hardware, or randomness. For this issue, we place it in the \texttt{Failure to replicate} category. In this paper, we followed three steps to categorize issues in the open-source AI repositories. These steps are described as follows:

\begin{enumerate}[label=(\roman*)]
	\item \textit{Preparation}: We randomly pick 384 issues\footnote{It is statistically representative sample size using a popular sample size calculator (\url{https://www.surveysystem.com/sscalc.htm}) with a confidence level of 99\% and a confidence interval of 10.} from all the collected issues, then we import these issues into a Notion database and create a card for each issue.
	\item \textit{Execution}: Two annotators discuss with each other how to categorize the cards into meaningful groups. We adopt the \textit{open card} sorting method, which lets the groups emerge and evolve during the sorting process. In this step, the two authors are encouraged to create groups. This step takes about seven hours.
	\item \textit{Analysis}: Two annotators who conduct card sorting, as well as a third annotator, discuss together how to merge groups of relevant topics into more general categories. This step takes approximately 1.5 hours.
\end{enumerate}

\pgfmathsetmacro{\MD}{10} 
\pgfmathsetmacro{\MM}{9} 
\pgfmathsetmacro{\MCM}{9} 
\pgfmathsetmacro{\MCSUM}{int(\MD+\MM+\MCM)} 
\pgfmathsetmacro{\UCUI}{56} 
\pgfmathsetmacro{\UDI}{12} 
\pgfmathsetmacro{\UEC}{7} 
\pgfmathsetmacro{\UISUM}{int(\UCUI+\UDI+\UEC)} 
\pgfmathsetmacro{\DM}{10} 
\pgfmathsetmacro{\EX}{42} 
\pgfmathsetmacro{\RE}{89} 
\pgfmathsetmacro{\DI}{35} 
\pgfmathsetmacro{\FTR}{13} 
\pgfmathsetmacro{\EN}{16} 
\pgfmathsetmacro{\AB}{50} 
\pgfmathsetmacro{\PCM}{5} 
\pgfmathsetmacro{\SI}{4} 
\pgfmathsetmacro{\RI}{15} 
\pgfmathsetmacro{\O}{2} 
\pgfmathsetmacro{\SUM}{int(\MCSUM+\UISUM+\DM+\EX+\RE+\DI+\FTR+\EN+\AB+\PCM+\SI+\RI+\O)} 

\pgfmathsetmacro{\MCSUMpercentage}{\MCSUM/\SUM*100}
\pgfmathsetmacro{\MDpercentage}{\MD/\SUM*100}
\pgfmathsetmacro{\MMpercentage}{\MM/\SUM*100}
\pgfmathsetmacro{\MCMpercentage}{\MCM/\SUM*100}
\pgfmathsetmacro{\UISUMpercentage}{\UISUM/\SUM*100}
\pgfmathsetmacro{\UCUIpercentage}{\UCUI/\SUM*100}
\pgfmathsetmacro{\UDIpercentage}{\UDI/\SUM*100}
\pgfmathsetmacro{\UECpercentage}{\UEC/\SUM*100}
\pgfmathsetmacro{\DMpercentage}{\DM/\SUM*100}
\pgfmathsetmacro{\EXpercentage}{\EX/\SUM*100}
\pgfmathsetmacro{\REpercentage}{\RE/\SUM*100}
\pgfmathsetmacro{\DIpercentage}{\DI/\SUM*100}
\pgfmathsetmacro{\FTRpercentage}{\FTR/\SUM*100}
\pgfmathsetmacro{\ENpercentage}{\EN/\SUM*100}
\pgfmathsetmacro{\ABpercentage}{\AB/\SUM*100}
\pgfmathsetmacro{\PCMpercentage}{\PCM/\SUM*100}
\pgfmathsetmacro{\SIpercentage}{\SI/\SUM*100}
\pgfmathsetmacro{\RIpercentage}{\RI/\SUM*100}
\pgfmathsetmacro{\Opercentage}{\O/\SUM*100}

\begin{table}[]
	\centering
	\caption{The distribution of issues of different categorizes.}
	\begin{tabular}{|c|ll|c|r|}
		\hline
		ID                 & \multicolumn{2}{l|}{Category}                  & No. & Rate \\ \hline
		\multirow{4}{*}{1} & \multicolumn{2}{l|}{Missing Content}           & $\MCSUM$     &  $\pgfmathprintnumber[fixed, fixed zerofill, precision=2]{\MCSUMpercentage}\%$          \\
						   &  & $\cdot$ Missing Dataset                     & $\MD$      &  $\pgfmathprintnumber[fixed, fixed zerofill, precision=2]{\MDpercentage}\%$           \\
						   &        & $\cdot$ Missing Model                         & $\MM$     &  $\pgfmathprintnumber[fixed, fixed zerofill, precision=2]{\MMpercentage}\%$          \\
						   &        & $\cdot$ Missing Code/Module                   & $\MCM$      & $\pgfmathprintnumber[fixed, fixed zerofill, precision=2]{\MCMpercentage}\%$           \\
		\multirow{4}{*}{2} & \multicolumn{2}{l|}{Unclear Instructions}      & $\UISUM$     &  $\pgfmathprintnumber[fixed, fixed zerofill, precision=2]{\UISUMpercentage}\%$          \\
						   &        & $\cdot$ Unclear Code Usage Instruction        & $\UCUI$     &  $\pgfmathprintnumber[fixed, fixed zerofill, precision=2]{\UCUIpercentage}\%$          \\
						   &        & $\cdot$ Unclear Data Information              & $\UDI$     &  $\pgfmathprintnumber[fixed, fixed zerofill, precision=2]{\UDIpercentage}\%$          \\
						   &        & $\cdot$ Unclear Environment Configuration     & $\UEC$      &  $\pgfmathprintnumber[fixed, fixed zerofill, precision=2]{\UECpercentage}\%$          \\
		3                  & \multicolumn{2}{l|}{Discuss Methodology}       & $\DM$     & $\pgfmathprintnumber[fixed, fixed zerofill, precision=2]{\DMpercentage}\%$           \\
		4                  & \multicolumn{2}{l|}{Extension}                 & $\EX$     &  $\pgfmathprintnumber[fixed, fixed zerofill, precision=2]{\EXpercentage}\%$          \\
		5                  & \multicolumn{2}{l|}{Runtime Error}             & $\RE$     &  $\pgfmathprintnumber[fixed, fixed zerofill, precision=2]{\REpercentage}\%$          \\
		6                  & \multicolumn{2}{l|}{Discuss Implementation}    & $\DI$     &  $\pgfmathprintnumber[fixed, fixed zerofill, precision=2]{\DIpercentage}\%$          \\
		7                  & \multicolumn{2}{l|}{Fail to Replicate}         & $\FTR$     &  $\pgfmathprintnumber[fixed, fixed zerofill, precision=2]{\FTRpercentage}\%$          \\
		8                  & \multicolumn{2}{l|}{Enhancement}               & $\EN$     &  $\pgfmathprintnumber[fixed, fixed zerofill, precision=2]{\ENpercentage}\%$          \\
		9                  & \multicolumn{2}{l|}{Abnormal Behavior}        & $\AB$     &  $\pgfmathprintnumber[fixed, fixed zerofill, precision=2]{\ABpercentage}\%$          \\
		10                 & \multicolumn{2}{l|}{Paper-Code Misalignment}   & $\PCM$      &  $\pgfmathprintnumber[fixed, fixed zerofill, precision=2]{\PCMpercentage}\%$          \\
		11                 & \multicolumn{2}{l|}{Supplementary Information} & $\SI$      &  $\pgfmathprintnumber[fixed, fixed zerofill, precision=2]{\SIpercentage}\%$          \\
		12                 & \multicolumn{2}{l|}{Re-implementation}         & $\RI$     &  $\pgfmathprintnumber[fixed, fixed zerofill, precision=2]{\RIpercentage}\%$          \\
		13                 & \multicolumn{2}{l|}{Others}                    & $\O$      &  $\pgfmathprintnumber[fixed, fixed zerofill, precision=2]{\Opercentage}\%$          \\ \hline
		\end{tabular}
	\label{tab:categories}
	\end{table}

\vspace{0.2cm}
\noindent \textbf{Categorization Results.}
We obtain 13 categories; two of them are further divided into three subcategories. Table~\ref{tab:categories} shows the categories and subcategories in the sampled issues. 
In our taxonomy, each issue is assigned to one category, which reflects the primary reason for raising this issue.
We discuss the details of these categories as follows:
\begin{itemize}
	\item \texttt{Missing Content}: 
Users raise issues as they are unable to find the contents used to employ open-source AI repositories. There are 28 issues that fall into this category, which is subsequently divided into three different subcategories, such as \textit{missing dataset} (10 issues), \textit{missing model} (9 issues), and \textit{missing code/module} (9 issues). For example, the issue\footnote{\url{https://github.com/pathak22/context-encoder/issues/24}} titled ``where to download the Paris StreetView Dataset'' requests the Paris StreetView dataset for running the Context Encoders model to generate the contents of an arbitrary image~\cite{pathak2016context}.
	\item \texttt{Unclear Instructions}: 
 The category is mentioned when users are unable to comprehend the instructions used to employ open-source AI repositories. This category contains 75 issues (19.53\%), which is the second-largest category. We define three subcategories based on information from the instructions. The first subcategory is \textit{unclear environment configuration}, e.g., the version of a programming language library in an AI repository.\footnote{https://github.com/ShichenLiu/CondenseNet/issues/22} The second subcategory is \textit{unclear data information}, e.g., how to split a dataset into training and testing sets. The third subcategory is \textit{unclear code usage instruction}, e.g., hyperparameters used to train AI models. 
	\item \texttt{Discuss Methodology}: 
 This category, which includes ten issues, focuses on discussing the methodology of AI models. For example, an issue\footnote{https://github.com/irwinherrmann/stochastic-gates/issues/1} asks the input size of an AI model.
	\item \texttt{Extension}: 
 Users raise their issues to ask AI repositories' owners whether AI models can be extended beyond their original scope. For example, users may want to apply AI models to a new dataset or do some modifications to these models.
	\item \texttt{Runtime Error}: 
 We define this category for users who encounter runtime errors when using AI repositories. The category includes 89 issues (23.18\%) which is the largest population of all categories. There are various reasons for the runtime errors, e.g., the users may not have the required packages installed, the users employ the wrong path, or the hardware is incompatible. We find that it is challenging to further divide these issues into subcategories based on the root cause of the runtime errors. 
	It is difficult to further divide these issues into subcategories based on the cause of the runtime errors. 
	\item \texttt{Discuss Implementation}: 
 This category aims to discuss the implementation details of AI repositories. For example, users may ask whether we should apply dropout techniques to AI models. Some users may also require the AI repositories' owners to clarify the details of their implementation.
	\item \texttt{Failure to Replicate}: 
 Users often submit their issues as they fail to employ an AI repository to replicate results reported in its corresponding paper. We put these issues in this category. 
	\item \texttt{Enhancement}: 
 Issues in this category aim to improve the quality of AI repositories, e.g., by adding new features, fixing bugs, or updating libraries. 
	\item \texttt{Abnormal Behaviour}: 
 This category focuses on users who may encounter performance issues, e.g., the training time of AI models takes too long.\footnote{https://github.com/open-mmlab/mmdetection/issues/5237}
	\item \texttt{Paper-Code Misalignment}: 
Issues related to this category focus on the difference between the code in AI repositories and the descriptions of their corresponding papers. For example, users may ask why the dimensions of hidden layers in an AI repository differ from the descriptions on its paper.\footnote{https://github.com/timmeinhardt/trackformer/issues/58} 
	\item \texttt{Supplementary Information}: 
 These issues request supplementary information for AI repositories, e.g., slides, documents, and videos of AI models.
	\item \texttt{Re-implementation}: 
 Users aim to re-implement AI repositories by employing other programming languages; hence, they may raise issues to discuss with the AI repositories' owners and other developers. 
	\item \texttt{Others}: 
 The remaining issues that are unable to categorize into any of the above categories. For example, some issues have insufficient information to be categorized.\footnote{\url{https://github.com/matterport/Mask_RCNN/issues/2046}} 
\end{itemize}

\begin{tcolorbox}[
	colback=green!10,
	colframe=green!50!black,
	colbacktitle={red!50!white},
	coltitle=white,
	fonttitle=\sffamily\bfseries\large,
	title=Answer to RQ1,
	center title,
	]
	We categorize the issues in open-source AI repositories into 13 classes, including: Missing Content, Unclear Instructions, Discuss Methodology, Extension, Runtime Error, Discuss Implementation, Failure to Replicate, Enhancement, Abnormal Behaviour, Paper-Code Misalignment, Supplementary Information, Re-implementation, and Others.
  \end{tcolorbox}

\subsection{RQ2. How are the issues managed and addressed in open-source AI repositories?}

\noindent \textbf{Motivation.}
The issues raised by users in open-source AI repositories serve as an important indicator of problems encountered during usage and also serve as a means of communication with repository maintainers. Effective management and addressing of these issues can greatly impact the user experience and the overall quality of the AI repositories. As the first research question highlights, users raise issues for a variety of reasons, such as missing content, unclear instructions, or runtime errors. Ignoring these issues can prevent users from successfully replicating the results reported in the repositories. While some issues can be easily resolved by users, such as installing required packages, other issues such as runtime errors or unclear instructions require better management and addressing in order for users to effectively employ the AI repositories for building their software applications. Considering the rapid growth of the AI field, it is crucial to investigate these issues to aid maintainers in effectively managing open-source AI repositories.

\begin{figure}[!t]
	\begin{minipage}[t]{0.333333333\linewidth}
	\centering
	\includegraphics[scale=0.35]{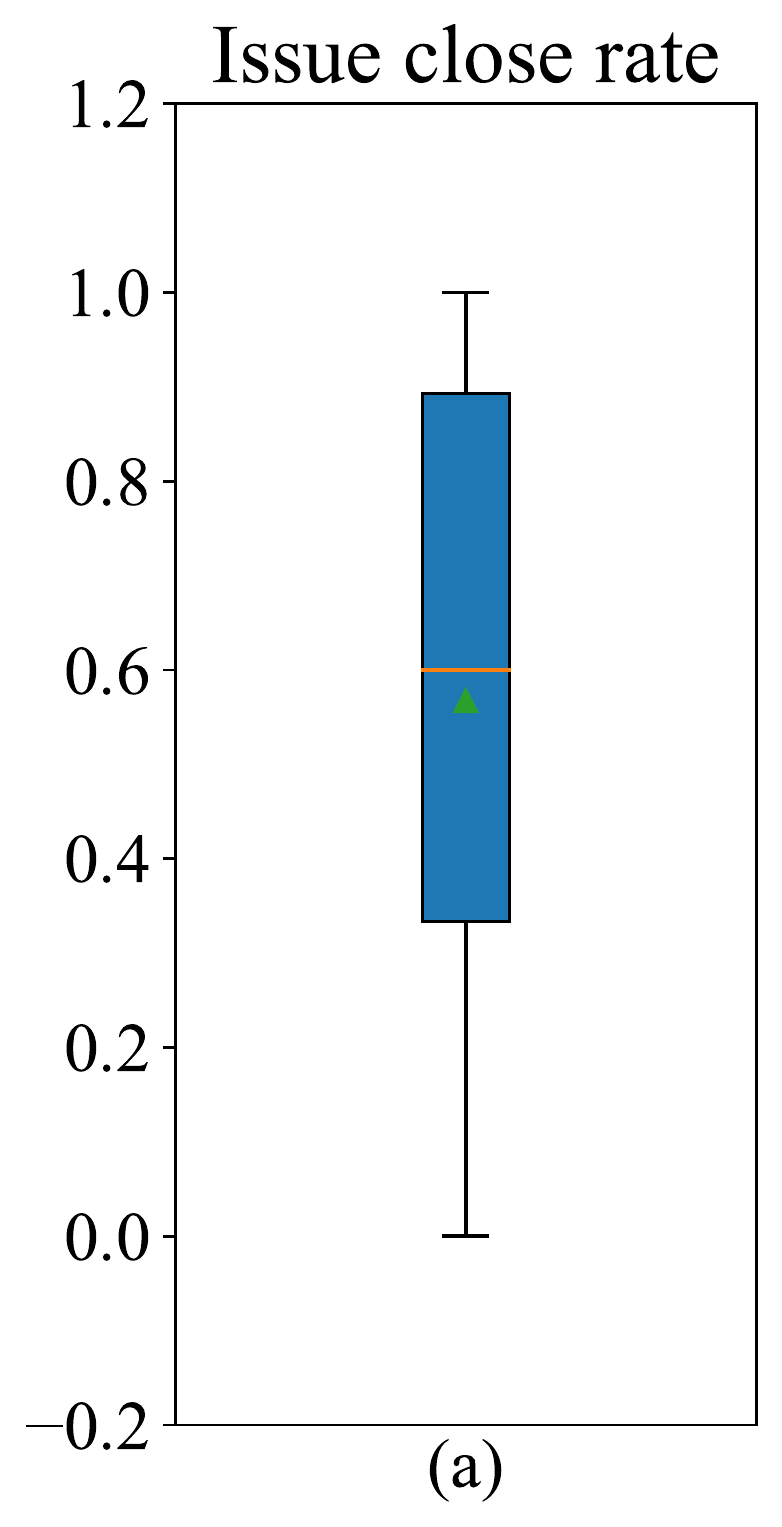}
	\label{fig:close-rate-dist}
	\end{minipage}%
	\begin{minipage}[t]{0.333333333\linewidth}
	\centering
	\includegraphics[scale=0.35]{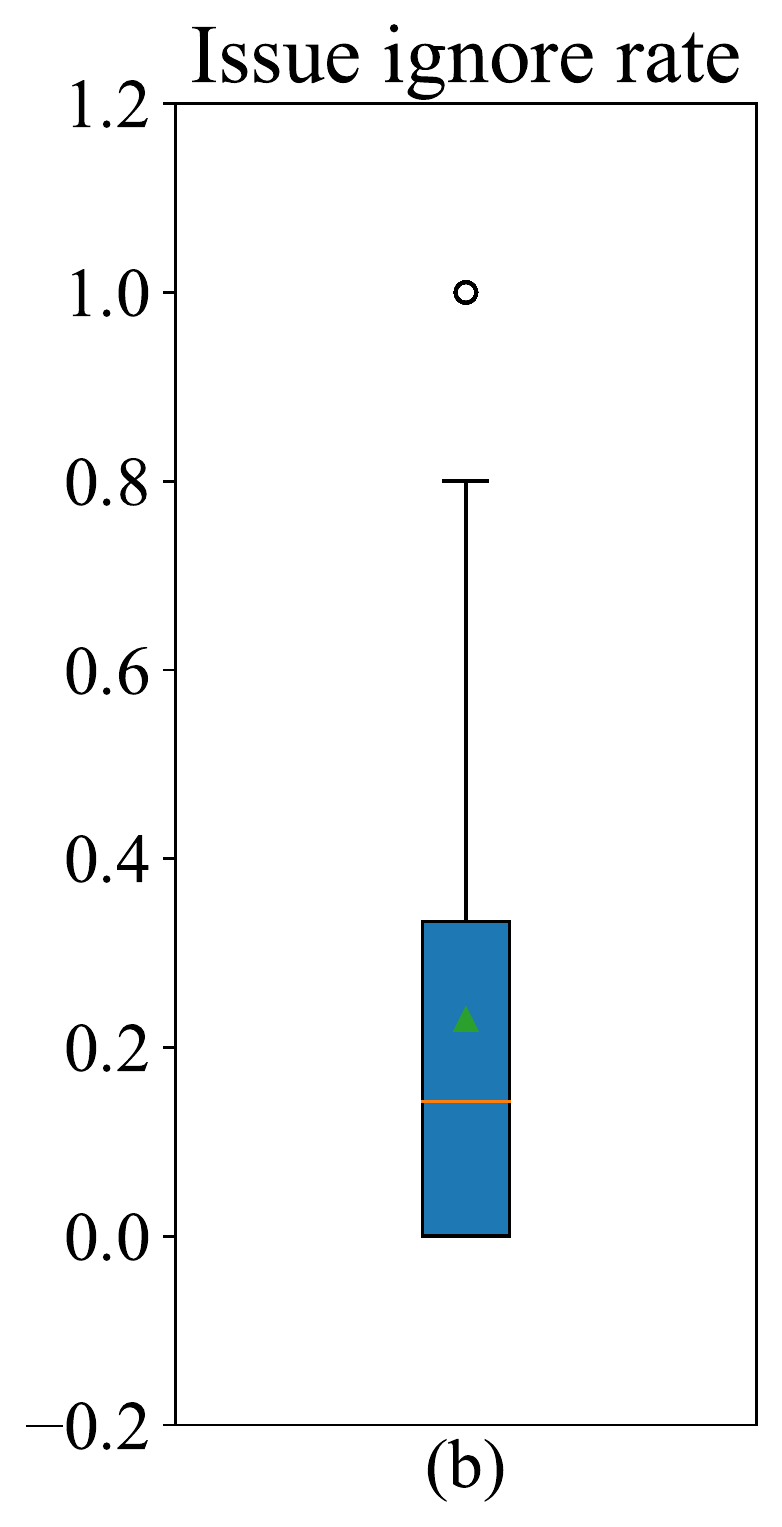}
	\label{fig:ignore-rate-dist}
	\end{minipage}%
	\begin{minipage}[t]{0.333333333\linewidth}
	\centering
	\includegraphics[scale=0.35]{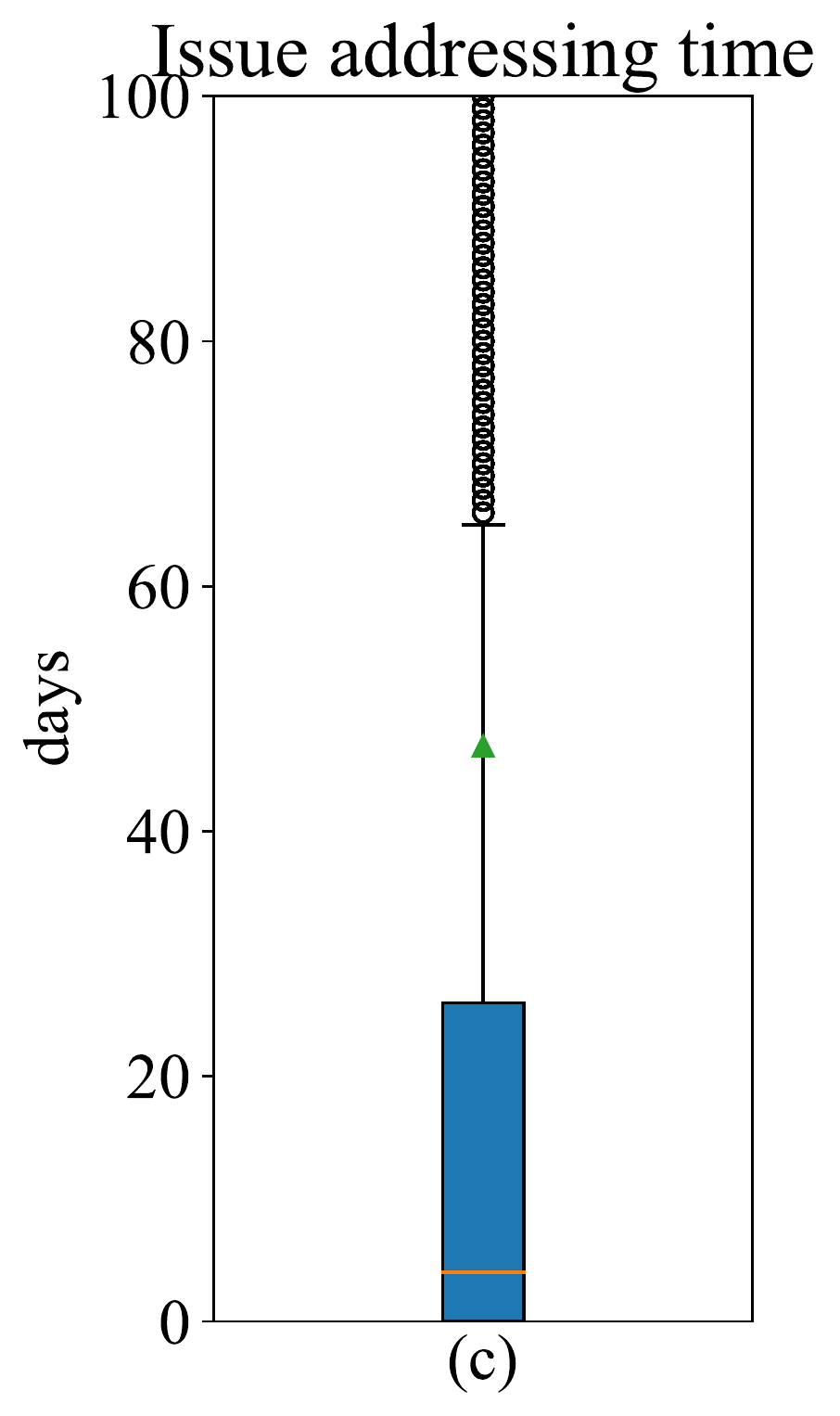}
	\label{fig:address-time-dist}
	\end{minipage}
	\caption{The distribution of closed issue rates (a), addressing time (b), and ignored issue rates (c) in investigated repositories.}
	\end{figure}

\vspace{0.2cm}
\noindent \textbf{How are the issues addressed?}
We aim to understand how to address the issues present in open-source AI repositories. These issues are categorized into two groups, \textit{open} and \textit{closed} issues. Open issues are those that have yet to be addressed, while closed issues are those that have been resolved. Utilizing the GitHub REST API, we collected 24,953 issues from open-source AI repositories. Of these issues, 8,102 (32.47\%) are open issues, and 16,851 (67.53\%) are closed issues. As shown in Figure~\ref{fig:close-rate-dist} (a), the distribution of the \textit{closed issue rate} (CIR) for each repository reveals that 35.76\% of repositories have closed all of their issues. However, it is worth noting that these repositories with a CIR higher than 0.9 have an average of 83.83 issues, indicating that these repositories have a larger number of issues in total.

We discover that 40\% of closed issues are ``\textit{self-closed},'' indicating that the individuals who raised the issues are able to find solutions to their problems through discussion. Additionally, we identify another type of issue referred to as \textit{ignored issues}. These are issues that only the individuals who raised them are interested in and typically receive no response from other users or developers. These ignored issues comprise 11.79\% of all issues. We also define the \textit{ignored issue rate} as the ratio of ignored issues for each repository. This metric allows us to evaluate the level of active engagement by repository maintainers in addressing issues. The results of the ignored issue rate are presented in Figure~\ref{fig:close-rate-dist} (b).



Additionally, we investigate the addressing time of issues, i.e., the time between the creation of an issue and its closure. We found that the average addressing time and standard deviation are 46.94 days and 135.34 days, respectively, in the 16,851 closed issues. However, the median addressing time is only four days, while the longest addressing time is over two years. Figure~\ref{fig:close-rate-dist} (c) illustrates the distribution of the addressing time. We analyze what factors affect the addressing time by considering two features at the repository level: (1) the number of contributors and (2) the total number of issues. We hypothesize that more contributors can reduce the addressing time. However, Spearman's rank correlation~\cite{sedgwick2014spearman} shows that the relationship between the number of contributors and the addressing time is not statistically significant ($p$-value $>0.05$). We also hypothesize that more issues can increase the average addressing time; intuitively, more issues mean more workload on the maintainers. Spearman's rank correlation validates this hypothesis by finding a weak correlation between the number of issues and average addressing time ($p$-value $<0.01$). 

\vspace{0.2cm}
\noindent \textbf{How are the issues managed?}
GitHub provides several methods to help maintainers manage issues. 
We consider two methods: (1) using labels to tag issues and (2) assigning issues to specific people to address. In the remaining part of this section, we use the terms \texttt{labeling} and \texttt{assigning} to refer to these two methods, respectively. However, our results show that the two methods are not widely adopted to manage issues in open-source AI repositories. Only 7.81\% of the repositories use labels to categorize issues, and only 5.90\% of repositories assign issues to assignees. We find that 11.42\% of issues are tagged with labels and 16.45\% of issues have assignees, which suggests that \textit{repositories with more issues tend to use these two methods to manage issues more often}.


\begin{figure}[!t]
	\centering
	\includegraphics[width=1\linewidth]{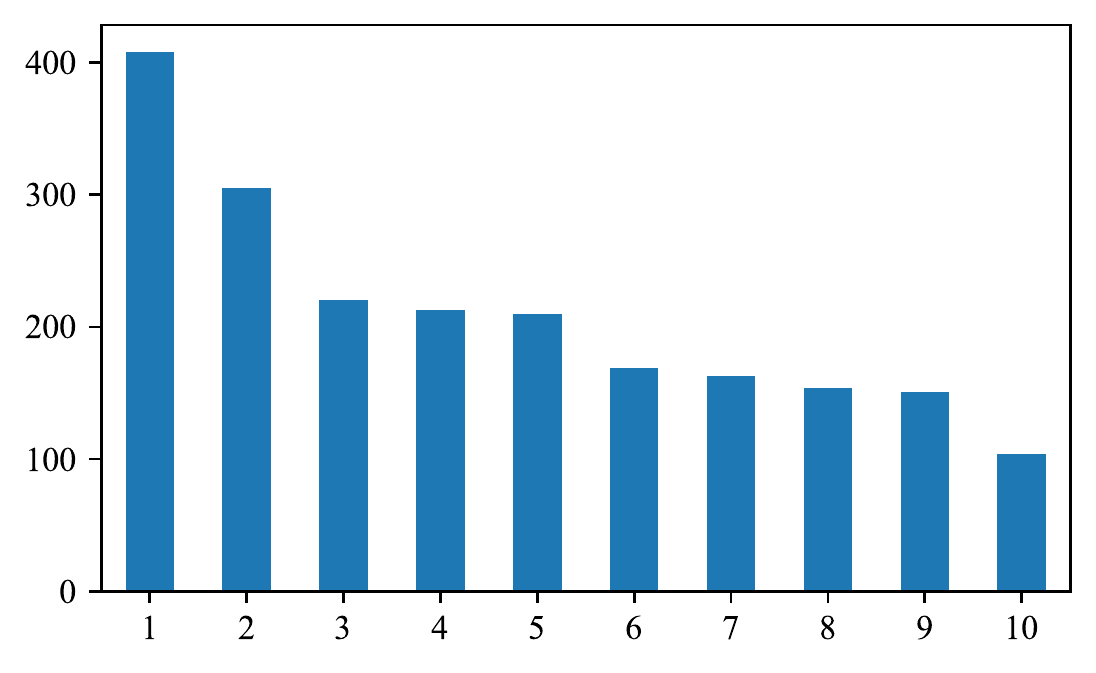}
	\caption{
		The distribution of the top 10 most frequently used labels. The numbers represent the labels, where 
		1: \texttt{bug}, 2: \texttt{enhancement}, 3: \texttt{awaiting response}, 4: \texttt{community help wanted}, 5: \texttt{help wanted}, 6: \texttt{good first issue}, 7: \texttt{community discussion}, 8: \texttt{reimplementation}, 9: \texttt{question}, 10: \texttt{feature request}.
		}
	\label{fig:label-occurences}
\end{figure}

We analyze the use of \texttt{labeling}. GitHub provides a set of nine default labels, such as \textit{bug}, \textit{duplicate}, \textit{enhancement}, \textit{help wanted}, \textit{invalid}, \textit{question}, \textit{wontfix}, and \textit{good first issue}. One sign of actively managing issues is to define new labels beyond just using the default ones. We find that among all the repositories that use labels, 14.41\% of them describe new labels. Including the default labels, we discover 109 unique labels being used in all the repositories. We then count the occurrences of each label. Table~\ref{fig:label-occurences} shows the ranked list of the most frequently used default labels. 


\begin{tcolorbox}[
	colback=green!10,
	colframe=green!50!black,
	colbacktitle={red!50!white},
	coltitle=white,
	fonttitle=\sffamily\bfseries\large,
	title=Answer to RQ2,
	center title,
	]
	In the investigated open-source AI repositories, 67.53\% of issues are closed, and 50\% of issues are closed within four days. Only 7.81\% of the repositories use labels to categorize issues, and only 5.9\% of them assign issues to specific assignees.
	We find that repositories with more contributors do not necessarily address issues faster, but repositories with more issues tend to take a longer time to close issues.
  \end{tcolorbox}

\subsection{RQ3. What are the relationships between different features and the closure of issues in open-source AI repositories?}

\noindent \textbf{Motivation.}
This research question aims to explore the relationships between the various features of issues and the closure of those issues. The answer to this research question can provide practical suggestions for both repository maintainers and issue raisers to improve the management and resolution of issues in open-source AI repositories.

\vspace{0.2cm}
\noindent \textbf{Investigated Features.}
We study the relationships between nine features and the closure of issues. The first eight features are related to the issues and the 9$^{th}$ feature, i.e., \texttt{num-contributors} is related to the repositories. We present the nine features as follows: 

\begin{itemize}
	\item \texttt{has-label}: Whether an issue is labeled. In GitHub, anyone with triage access to a repository can tag issues with labels. A labeled issue usually means that it has been noticed and read by the maintainers and, as a result, it may be more likely to be addressed.
	\item \texttt{has-assignee}: 
    Whether an issue is assigned. Users who have a write access permission to a repository can assign issues to developers. If the issue is assigned, the assignee is responsible for addressing the issue, which may lead to a successful resolution.
	\item \texttt{title-length}: The length of the issue title. An informative title may help the repository maintainers better understand the issue at a glance. Usually, a longer title carries more information that can help catch the maintainers' attention, leading to a faster solution.
	\item \texttt{body-length}: The length of the issue body. Similarly, an informative description can help the repository maintainers better understand the issue. In some guidelines, the issue raisers are encouraged to provide a detailed description of the issue, which can help the maintainers better understand the issue.
	\item \texttt{sentiment}: The sentiment of the issue description. If an issue is impolite, e.g., includes negative sentiment, a repository maintainer may be reluctant to address it.
	\item \texttt{is-English}: Whether the issue is written in English. We noticed that 5.94\% of issues are written in a language other than English, e.g., Chinese, leading the repository maintainers to barely understand these issues to provide their solutions.
	\item \texttt{has-code}: Whether the issue description contains code blocks. Code blocks are useful for the repository maintainers to understand the problem. For example, issue raisers may include code snippets corresponding to error tracks in the issue description. 
	\item \texttt{has-url}: Whether the issue description contains URLs. GitHub uses Markdown syntax to format the issue description. An URL can be used to refer to another issue, a website, or an image. This information may help repository maintainers better understand the issue.
	\item \texttt{num-contributors}: The number of contributors in the repository. Intuitively, the more contributors in a repository, the more likely the issue is to be addressed.
\end{itemize}

\vspace{0.2cm}
\noindent \textbf{Feature Measurement and Statistical Test.}
We obtain issues by querying the GitHub API. We compute the length of the issue title and body by converting them into Python strings and using the results returned by the \texttt{len()} function. Following the previous work~\cite{github-disc}, we employ Senti4SD~\cite{Senti4SD}, a sentiment analysis tool specifically tuned for software engineering tasks, to compute the sentiment of the issue description. Senti4SD produces three sentiment scores: \textit{positive}, \textit{negative}, and \textit{neutral}. As we train Senti4SD on texts written in English, we discard the issues that contain non-English characters. For issues that contain code snippets and URLs that are not relevant to analyzing the sentiment of the issue description, we remove them using regular expressions.

We first tokenize the issue title and body into characters. We then check whether the issue contains characters that are in the American Standard Code for Information Interchange\footnote{\url{https://en.wikipedia.org/wiki/ASCII}} (ASCII). If the issue contains non-ASCII characters, we consider it \texttt{non-English}.
The GitHub issues use Markdown syntax, which allows users to format their issue description using code blocks (i.e., code snippet in \texttt{```} and \texttt{'''}) and URLs (i.e., \texttt{[text](URL)}).
We employ regular expressions to detect whether the issue description contains code blocks or URLs using Markdown syntax.

We group the issues into two groups: \textit{open} and \textit{closed} issues. For each issue, we measure its eight features, such as \texttt{has-label}, \texttt{has-assignee}, or \texttt{title-length}. The number of contributors in a repository is obtained by querying the GitHub API to fetch metadata about the repository. Then, we adopt the Wilcoxon rank-sum test~\cite{Wilcoxon} to compute the significance of the differences between the feature values of open and closed issues. We also estimate the Cohen's effect sizes $\delta$ of the difference between the open and closed issues. According to the guideline for using the effect size, when $|\delta|$ is less than 0.2, between 0.2 and 0.5, between 0.5 and 0.8, and larger than 0.8, the effect size is considered negligible, small, medium, and large, respectively.

\begin{table}[!t]
	\centering
	\caption{The significance of the differences between the feature values of open and closed issues.}
	\begin{tabular}{lcc}
		\hline
		Feature          & $p$-value & Cohen's $\delta$ \\ \hline
		\texttt{has-label}        & $< 0.01$ & 0.26 (S)          \\
		\texttt{has-assignee}     & $< 0.01$ & 0.50 (M)          \\
		\texttt{title-length}     & $> 0.05$  & 0.02 (N)             \\
		\texttt{body-length}      & $< 0.01$  & 0.08 (N)         \\
		\texttt{sentiment}        & $> 0.05$  &  0.01 (N)         \\
		\texttt{is-Non-English}       &  $> 0.05$  & 0.05 (N)       \\
		\texttt{has-code}         &  $< 0.01$ &  0.14 (N)          \\
		\texttt{has-pics}         &  $> 0.05$  & 0.01 (N)          \\
		\texttt{num-contributors} &  $< 0.01$ & 0.54 (M)         \\ \hline
		\end{tabular}
	\label{tab:features}
	\end{table}

Table~\ref{tab:features} shows the results of the Wilcoxon rank-sum test and Cohen's effect sizes $\delta$ for the nine features. Five features, such as \texttt{has-label}, \texttt{has-assignee}, \texttt{body-length}, \texttt{has-code}, and \texttt{num-contributors}, show a statistically significant difference between the open and closed issues. The \texttt{num-contributors} feature also has a medium effect size (M), indicating that the more contributors who join the repository, the better we can address the issues. The remaining features, i.e., \texttt{title-length}, \texttt{sentiment}, \texttt{is-English}, and \texttt{has-url}, indicate that there is no significant difference between the open and closed issues. 

We classify the issue-related features into two groups, i.e., features related to the practices of raising issues and features related to the practices of handling issues. Our results show that issues are more likely to be closed if they have assignees. Moreover, issues with labels are more likely to be closed than those without labels. The effect sizes of \texttt{has-label} and \texttt{has-assignee} are small and medium, respectively. 
The two features, i.e., \texttt{body-length} and \texttt{has-code}, are related to how users raise issues. Although their effect sizes are smaller than that of \texttt{has-label} and \texttt{has-assignee}, they still demonstrate statistically significant differences.


\begin{tcolorbox}[
	colback=green!10,
	colframe=green!50!black,
	colbacktitle={red!50!white},
	coltitle=white,
	fonttitle=\sffamily\bfseries\large,
	title=Answer to RQ3,
	center title,
	]
	Our results show that issues are more likely to be closed if they have assignees. Moreover, issues with labels are more likely to be closed than those without labels. Besides, issues with more details (e.g., longer descriptions and code blocks) have a higher chance of being closed.
  \end{tcolorbox}

\section{Discussion}
\label{sec:discussion}

\begin{table}[]
    \centering
    \caption{The difference between industry and academic repositories.}
    \begin{tabular}{lcccc}
    \hline
    Class            & Academic & Industry & $p$-value          & $\delta$   \\ \hline
    No. Issues       & 22.57    & 257.62   & \textless{}0.01    & Small      \\
    No. Stars        & 232.82   & 2749.10  & \textless{}0.01    & Medium     \\
    No. Contributors & 2.25     & 41.06    & \textless{}0.01    & Small      \\
    Address Time     & 45.47    & 47.90    & \textless{}0.01    & Negligible \\
    Close Rate     & 57\%     & 58\%     & \textgreater{}0.05 & Negligible \\
    Assign Rate      & 1.35\%  & 5.15\%  &  \textless{}0.05  & Small   \\
    Label Rate       & 2.45\%    & 6.45\%  &  \textgreater{}0.05      &  Small \\ \hline
    \end{tabular}
    \label{tab:difference}
\end{table}


\subsection{Industry vs. Academic Repositories}

Developers from industry and academia work closely to promote the development of AI systems together. 
On the one side, industry practitioners also use academic AI repositories to build their products.
On the other side, academic researchers use popular AI frameworks originating from industry to create their models.
We are also interested in the difference between issues in the industry and academic repositories. 
Similar to the process in~\cite{NICHE}, we ask two annotators to manually split collected repositories into two types: repositories owned by industry and those owned by academia. 
The inter-annotator agreement value is 0.62, indicating substantial agreement among our annotators~\cite{landis1977measurement}.
We invite another annotator to resolve the disagreement between the two annotators. 
In total, we have 63 repositories owned by industry and 513 repositories owned by academia. We compare the two groups of repositories from the perspectives of the number of issues, addressing time, close rate, label distribution, assigning rate, and labeling rate. We conduct a Wilcoxon rank-sum test to check whether the difference is statistically significant. Table~\ref{tab:difference} presents our results.

On average, each industry repository includes 257.62 issues, while each academic repository contains 22.57 issues. 
However, it does not necessarily indicate that industry repositories have low quality. The average number of stars in industry repositories (2749.10) is one order of magnitude greater than that in academic repositories (232.82), meaning that industry repositories attract much more attention and, therefore, the number of issues in these repositories is larger. Note that the ratio of the number of issues to the number of stars is close: 0.094 for industry repositories and 0.097 for academic ones.

The industry repositories have over 18 times more average contributors than the academic ones (41.06 vs. 2.25). The ratio of the number of issues to the number of contributors is 6.27 for industry repositories and 10.03 for academic repositories. However, a smaller issue-contributor ratio does not lead to faster issue-addressing time; industry repositories even have a slightly (but statistically significant) longer addressing time (47.90 days) than academic repositories (45.47 days). Besides, the issue close rate is also similar between the two types of repositories (0.58 and 0.57). One possible explanation is that the academic repositories are less active after the code is released, meaning that the contributors mainly focus on addressing the issues. 
However, industry repositories require their contributors to put more effort into both developing new features and addressing issues.

The two types of repositories also demonstrate different behaviors in using the issue management tools. Although both the labeling and assigning functions are not widely adopted in the open-source AI repositories, there are more issues in industry repositories that are labeled and assigned. The assign rates are 5.15\% and 1.35\% for industry and academic repositories, respectively; the difference is also statistically significant ($p\text{-value}<0.05$). The label rates are 6.45\% and 2.45\% for industry and academic repositories, respectively. However, the difference is not statistically significant ($p\text{-value}>0.05$). Overall, the industry repositories are more active in using the issue management features than the academic ones.

\subsection{Implications}
Based on the analysis of each research question, we summarize the implications of our study as the suggestions to open-source AI repository developers and users.

In this study, we find that the most prevalent issue among the open-source AI repositories is the \texttt{Runtime Error} category, which can be caused by a variety of factors related to other categories of issues, such as missing datasets or missing models. The second most common category of issues falls under the \texttt{Unclear Instructions}, which is primarily caused by a lack of detailed descriptions. 
To address these issues and improve the quality of open-source AI repositories, we propose two suggestions for the maintainers: (1) ensure that all necessary datasets, models, and other files are provided to allow for accurate replication of results, and (2) provide clear and detailed instructions on how to use the repository, including information on dataset splitting, setting hyperparameters, and managing library dependencies.

Our experiment results show that issues are more likely to be closed if they have assignees and labels. 
Therefore, we recommend that repository maintainers actively leverage the features provided by GitHub to manage issues by (1) tagging issues with labels and (2) assigning specific people to be responsible for addressing the issues.
Besides, the length of issues descriptions and whether the issues have code blocks cause a statistically significant difference to the closure of issues. 
Based on the results, we make two suggestions to the issue raisers: (1) provide more details to precisely describe the issue and (2) use the Markdown syntax to formulate the issue description with code snippets.

\subsection{Threats to Validity}

\noindent \textbf{Threats to Internal Validity.}
We collect a list of repositories from the PaperWithCode platform, which is a community-driven platform where users can link papers with their corresponding repositories. However, there is a risk that some of these repositories may not be official implementations, which can be of low quality and attract less attention from users. To mitigate this threat, we only keep the repositories that are marked as ``official implementations.'' Additionally, we compare the repositories and paper information, such as owner and author information, to ensure that the repositories are official. This process is done manually, with two authors involved, and any disagreements are resolved with the involvement of another annotator.
Furthermore, we manually group the repositories by their ownership, such as industry and academia. However, considering that academia and industry work closely together in AI research, it may be difficult to unanimously agree on the ownership of a repository. In such cases, another annotator is involved in the process to lead the discussion and resolve any disagreements.

\vspace*{0.2cm}
\noindent \textbf{Threats to External Validity.}
Our study collects papers and repositories from the flagship AI conferences within a ten-year timeframe. However, it is important to note that there is a risk that the conclusions drawn from our study may not be generalizable to a broader range of open-source AI repositories. The field of AI research encompasses a wide range of subtopics, some of which are more popular than others. This leads to an imbalanced distribution of repositories tackling various topics in our dataset, which may affect the generalizability of our conclusions to other AI subtopics or interdisciplinary AI research.
Additionally, while we analyze open-source AI repositories hosted on GitHub, which is the most popular platform for open-source AI projects,  other platforms such as GitLab or BitBucket, which employ different issue trackers, such as JIRA, also contain open-source AI repositories. These platforms may have different issue management strategies, which could lead to diverse conclusions. 

AI is a fast-evolving field. Future work that replicates our study in a wider scope of open-source AI repositories, e.g., more venues, longer time periods, more AI subtopics, or other hosting platforms, can further validate and generalize our findings. We also make the replication package publicly available to facilitate such replication.

\section{Related Work}
\label{sec:related_work}

\subsection{Challenges and Practices in AI Development}

There has been a series of works investigating the challenges and best practices of various stages in the development of AI software, e.g., data collection, system designing, model testing and deployment, etc.

Whang and Lee~\cite{Whang} discussed the challenges of data collection for AI systems.
Amershi et al.~\cite{amershi2019SE4AI} conducted surveys and interviews to understand the process of developing AI software in Mircosoft teams. 
They organized the response from developers into a set of best practices. 
For example, in the data collection stage, they suggested reusing the data as much as possible to reduce duplicated effort.
Paleyes et al.~\cite{paleyes2021towards} provided suggestions on using flow-based programming to better discover and collect data for AI systems. 
Researchers also investigate the code smells~\cite{10.1145/3522664.3528620} and data smells~\cite{10.1145/3522664.3528621,10.1145/3522664.3528590} of AI software systems.
Serban et al.~\cite{10.1145/3382494.3410681} evaluated the adoption and effects of using conventional software engineering practices in AI software development.
Wan et al.~\cite{8812912} analyzed differences between the development of machine learning systems and the development of non-machine-learning systems, which derives recommendations for AI developers.

Song et al.~\cite{10.1145/3522664.3528596} explored the practices of testing machine learning software via conducting an interactive rapid review with industry practitioners.
Software engineering researchers designed a list of methods to test and improve different AI systems (e.g., speech recognition~\cite{asrdebugger,asdf-paper,crossasrpp}, reinforcement learning~\cite{MDPFuzz}, language models~\cite{checklist}, etc) from various perspectives beyond correctness, e.g., fairness~\cite{CFSA,biasfinder,biasrv,biasheal}, robustness~\cite{acsac2022gong,yang2022revisiting,monash-fairness}, security~\cite{advdoor,baffle}, etc. 
Shneiderman~\cite{shneiderman2020bridging} discussed the guidelines for addressing the ethics and reliability issues in human-centered AI systems.
Paleyes et al.~\cite{10.1145/3533378} presented a survey of case studies of the challenges and practices in deploying AI systems.
Fan et al.~\cite{fan2021makes} mined academic AI repositories and provided some suggestions to improve the quality and popularity of open-source AI repositories.

\subsection{Mining GitHub Issues}

Researchers have investigated the bugs in deep learning systems~\cite{10.1145/3377811.3380395,6405375}.
This paper mainly focuses on the issues raised in AI repositories, which according to our taxonomy in RQ1 covers contents beyond bugs, e.g., paper discussion. 
We also introduce the works on mining GitHub issues.

GitHub Issues~\cite{preston-werner_2009} is an issue-tracking system integrated into GitHub that allows users to track tasks, enhancements, and bugs~\cite{shaowei-icst} for repositories hosted on GitHub. A series of empirical studies have been established to analyze software repositories and their issues. In 2013, Bissyande et al.~\cite{6698918} conducted a large-scale investigation of issue trackers from tens of thousands of GitHub projects and obtained interesting findings, i.e., the correlation between the numbers of issue raisers and the addressing time.
\texttt{Ticket Tagger} was presented to automatically assign labels to issues in the GitHub issue tracker~\cite{KALLIS2021102598,8918993}. In this paper, we analyze the influence of labels in managing GitHub issues of AI repositories. Our results show that overall the label function is not frequently used in open-source AI repositories. Moreover, there is a significant difference in label usage distribution between open and closed issues. We believe tools for automatically tagging issues in AI repositories can boost the efficiency of issue resolution~\cite{izadi2022predicting,WANG2022106845}.
Recently, Hata et al.~\cite{hata2022github} analyzed the early usage of GitHub Discussion, a new feature of GitHub that allows users to discuss issues and topics. Their approach, i.e., the sentiment of the content, is also similar to our study. However, their paper does not focus on issue tracking systems.


\section{Conclusion and Future Work}
\label{sec:conclusion}

This paper presents the first systematic empirical study of issues in open-source AI repositories. Using the \texttt{PapersWithCode} platform, we collected 576 open-source AI repositories hosted on GitHub and 24,953 issues from these repositories. Through manual analysis of a representative sample of issues, we developed a taxonomy of 13 categories, with the two most common being \texttt{runtime error} (23.18\%) and \texttt{unclear instructions} (19.53\%). Our findings also showed that 67.5\% of issues were closed, with half of them closed within five days. Additionally, we discovered that GitHub issue management features were not widely adopted in open-source AI repositories, with only 7.81\% and 5.9\% of repositories using labels and assignees, respectively. We empirically show that employing GitHub issue management features and writing issues with detailed descriptions (e.g., using more code blocks) can help resolve issues. Our study provides recommendations for repository maintainers to improve the management and quality of open-source AI repositories. 

In the future, we intend to conduct surveys and interviews to uncover the reasons behind the lack of widespread adoption of issue management features in open-source AI repositories and to explore ways to increase their usage. 
Additionally, we will investigate the topics of discussion surrounding the development of AI systems, such as AI testing and data management, on platforms like StackOverflow and GitHub.

\vspace*{0.2cm}
\noindent \textbf{Reproducibility.}
The code and documentation, along with the obtained data, have been made open-source for reproducibility: \textbf{\url{https://github.com/soarsmu/Mining-AI-repos-issues}}.

\section*{Acknowledgment}
This research is supported by the Ministry of Education, Singapore under its Academic Research Fund Tier 3 (Award ID: MOET32020-0004). Any opinions, findings and conclusions or recommendations expressed in this material are those of the author(s) and do not reflect the views of the Ministry of Education, Singapore.

\bibliographystyle{IEEEtran}
\balance
\bibliography{../reference}

\end{document}